\documentclass[12pt]{article}
\usepackage{graphicx}

\usepackage{amsmath}
\usepackage{slashed}
\newcommand\pubnumber{SNSN-323-63}
\newcommand\pubdate{\today}

\def\institute{Institut Pluridisciplinaire Huber Curien\\
Strasbourg, France}
\def\Title#1{\begin{center} {\Large #1 } \end{center}}
\def\Author#1{\begin{center}{ \sc #1} \end{center}}
\def\Address#1{\begin{center}{ \it #1} \end{center}}

\newcommand\pubblock{\rightline{\begin{tabular}{l} \pubnumber\\
         \pubdate  \end{tabular}}}
\newenvironment{Abstract}{\begin{quotation}  }{\end{quotation}}
\newenvironment{Presented}{\begin{quotation} \begin{center} 
             PRESENTED AT\end{center}\bigskip 
      \begin{center}\begin{large}}{\end{large}\end{center} \end{quotation}}

\def\beq{\begin{equation}}
\def\eeq#1{\label{#1}\end{equation}}
\def\eeqn{\end{equation}}

\def\beqa{\begin{eqnarray}}
\def\eeqa#1{\label{#1}\end{eqnarray}}
\def\eeqan{\end{eqnarray}}

\let\bar=\overbar

\def\Dslash{\not{\hbox{\kern-4pt $D$}}}
\def\dslash{\not{\hbox{\kern-2pt $\del$}}}

\def\msb{{\bar{\ssstyle M \kern -1pt S}}}

\begin{document}
\begin{titlepage}
\pubblock

\vfill
\Title{Search for Dark Matter with top quarks}
\vfill
\Author{ Jeremy Andrea, \\on behalf of the ATLAS and CMS collaborations}
\Address{\institute}
\vfill
\begin{Abstract}
This proceeding presents searches for Dark Matter particles produced in association with top quarks at the LHC. The searches are performed by the ATLAS and CMS collaborations and various models and topologies are investigated. They are exploiting $t\bar{t}$ and single top experimental signatures by searching for an excess of missing transverse energy $\slashed{E}_T$. No signs of Dark Matter particles haven been observed  and limits on the models are set.
\end{Abstract}
\vfill
\begin{Presented}
$9^{th}$ International Workshop on Top Quark Physics\\
Olomouc, Czech Republic,  September 19--23, 2016
\end{Presented}
\vfill
\end{titlepage}
\def\thefootnote{\fnsymbol{footnote}}
\setcounter{footnote}{0}

\section{Introduction}

Astrophysical and cosmological observations seem to show evidence of the existence of a new type of particle, that can be observed only through gravitational effects. These new potential particles, named Dark Matter (DM) particles, should then be massive, should not carry an electric charge and should weakly interact with the common matter. They are called WIMP for Weakly Interactive Massive Particle.

DM particles can be searched for in multiple ways. First with cosmic sources  by looking for DM particles scattering or annihilation. But DM particles can also be produced at the LHC, either directly or through the decay of heavy new particles. DM particles are predicted by numerous BSM (Beyond the Standard Model) models, such as the R-parity conserving Supersymmetry.  However, following a more bottom-up approach, the theoretical framework of DM searches usually relies on simplified models. A review and discussions of these models can be found in \cite{DMforum}. In the following, I will present the searches for DM particles performed by the ATLAS\cite{atlas} and CMS collaborations\cite{cms}.

In this proceeding, searches for DM particles produced in association with top quarks are discussed.  The theoretical framework and the signatures are discussed in Sec.\ref{sec:theo}. The searches for DM in the $t\bar{t}$ channels and single top (monotop) channels are shown in Sections \ref{sec:ttbar} and \ref{sec:monotop}, respectively.  This proceeding will end with a short conclusion in Sec.\ref{sec:conclusion}.

\section{Theoretical framework}
\label{sec:theo}

Various models can predict DM candidates. While top-down approach usually gives the best sensitivity for a very specific model, it makes the search very dependent to assumptions. For this reason, a less model-dependent approach is followed. The usage of Effective Field Theory or simplified model is mainly based on the definition of the signature ($t\bar{t}+\slashed{E}_T$ or $t+ \slashed{E}_T$) and on the construction of the proper Lagrangian terms that allow for the production of the signatures of interest.
 
Several implementations of model and conventions are used by the ATLAS and CMS collaborations. A common effort for harmonizing the conventions, including theorists is performed within the Dark Matter Forum \cite{DMforum}. The idea behind is to harmonize the models so results from the ATLAS and CMS collaborations can be better compared. The experiments are in the process of converging toward very same models, but recent results are not yet fully comparable between the experiments.
 
\begin{figure}[htb]
\centering
\includegraphics[height=1.5in]{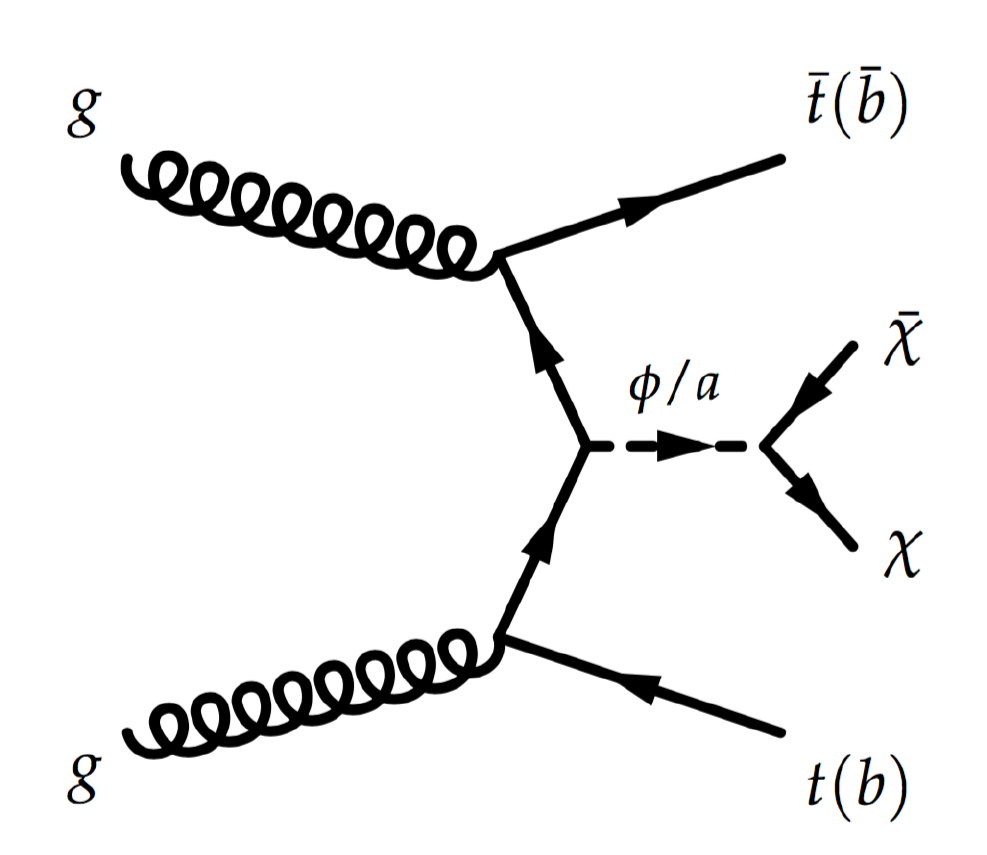}
\includegraphics[height=2.in]{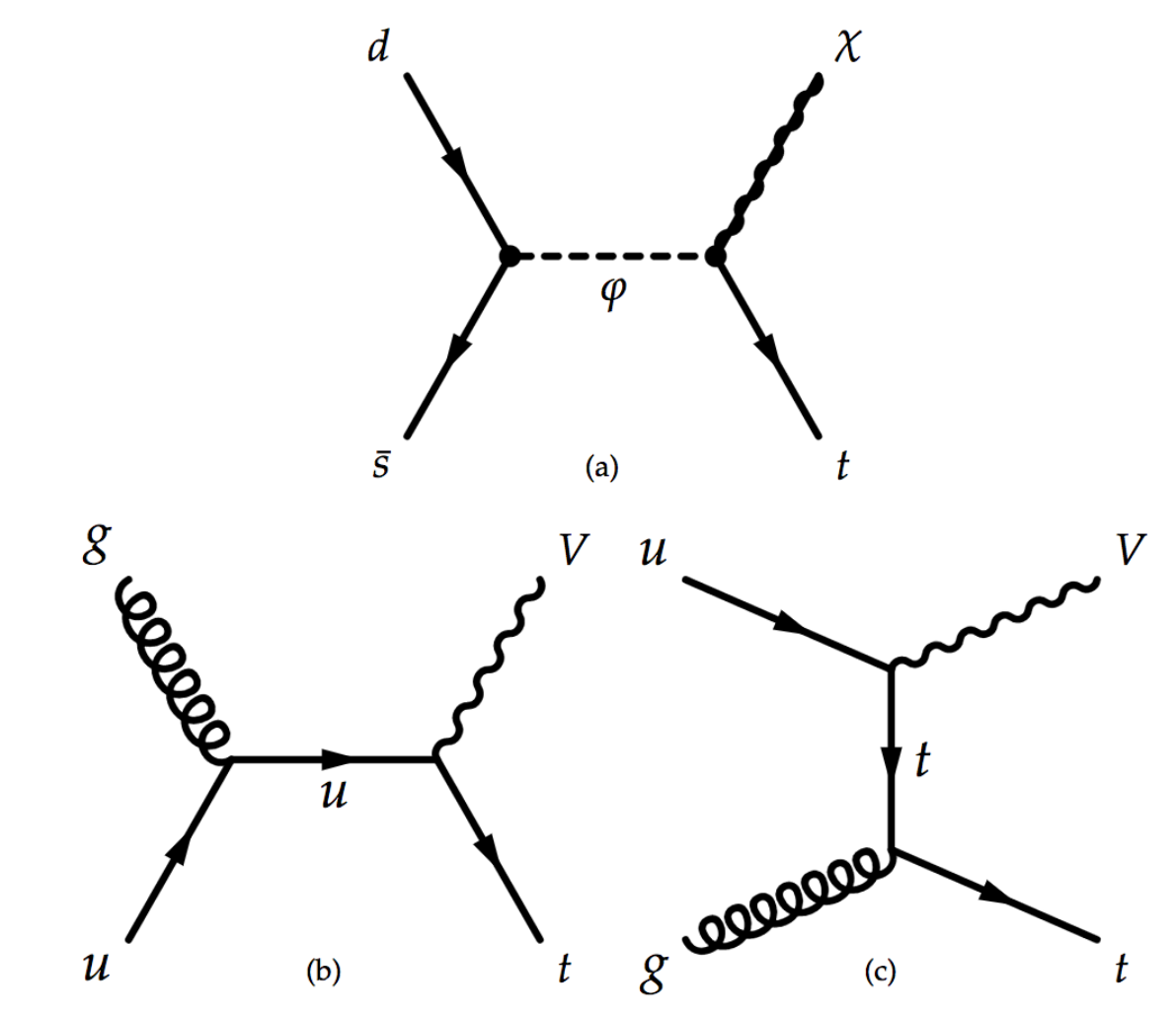}
\caption{Examples of Feynman diagrams for the production of top pairs and DM particles (left) and for the production of monotop (right) at the LHC.}
\label{figFeynm}
\end{figure}

Two main signatures are investigated. The first one presented in this document is the search for $t\bar{t}$ pairs produced with an extra source of $\slashed{E}_T$, arising from the presence of Dark Matter particles. The corresponding Feynman diagrams can be found on the left hand of Fig.\ref{figFeynm}. The Lagrangian terms that need to be added to the Standard Model (SM) Lagrangian allow to describe the production of new scalar ($\phi$) or pseudo scalar ($a$) particles that can decay into DM particles $\chi$. These terms are 

\begin{equation}
\mathcal{L} = g_\chi \phi \bar{\chi}\chi + \frac{\phi}{\sqrt{2}} \sum_q g_q^\phi y_q\bar{q}q + g_\chi a \bar{\chi} \gamma^5 \chi + \frac{ia}{\sqrt{2}}\sum_q g_q^a y_q \bar{q}\gamma^5q + h.c.,
\end{equation}

\noindent  with the couplings $g^\phi (g^a) = g_\chi = 3.5$ for the ATLAS results and  1 for the CMS results. The field $\chi$ is a new fermion being the DM candidate and the $y_q$ are the Yukawa couplings of the quark $q$.  The searches are performed by scanning over the values of the new particle masses : $m_\phi (m_a)$ and $m_\chi$. The benchmark points have been defined from the expected sensitivity at LHC and by considering a minimal change of kinematics.

DM particles can also be produced in association with a recoiling single top quark. This is the so called monotop signature, which is composed of a single top quark produced with larger $\slashed{E}_T$. In the current models, monotops can be described by either a new flavor changing neutral interaction (non-resonant) or via the production of a new scalar resonance $\phi$.  $V$ and $\chi$ are new vector or fermion with either long live times or  are decaying invisibly. The corresponding Feynman diagrams are shown in the three right plots of Fig.\ref{figFeynm}. The corresponding Lagrangian terms are :

\begin{equation}
\mathcal{L} = a_{res}\phi \bar{d^c}P_Rs + b_{res} \phi \bar{\chi}P_R t + a_{nonres} V_\mu \bar{u}\gamma^\mu P_R t + h.c.,
\end{equation}

\noindent with the couplings $a_{res}=b_{res}$ for ATLAS results while CMS results are produced with fixing $a_{res}$ and by considering a branching fraction of 100\% of the resonant particle into a top quark and $\chi$.

\section{Search for DM with $t\bar{t}$ signatures}
\label{sec:ttbar}

In this section are presented the ATLAS and CMS searches for DM particles produced in association with a pair of top quarks.

%\subsection{ATLAS searches for  $t\bar{t}$  + DM}

\begin{figure}[htb]
\centering
\includegraphics[height=2.in]{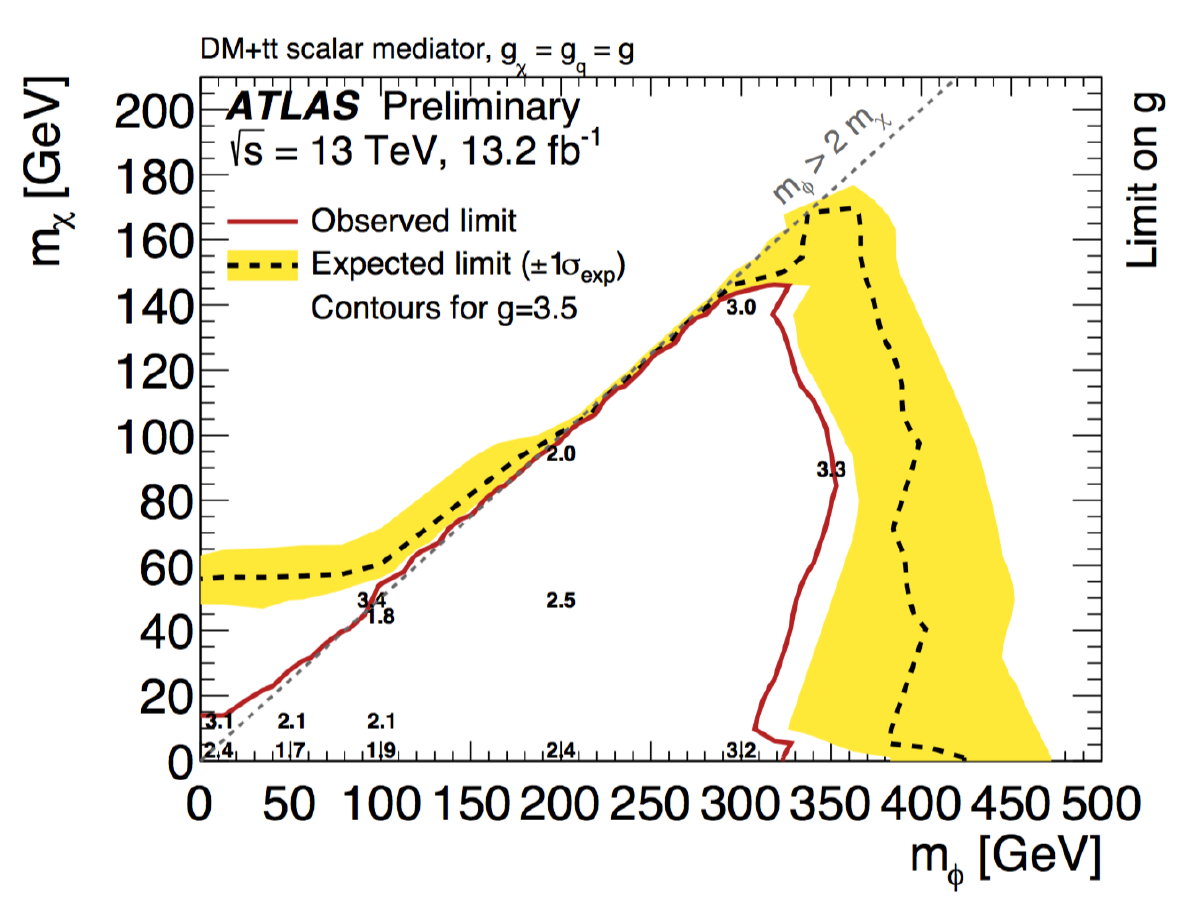}
\includegraphics[height=2.in]{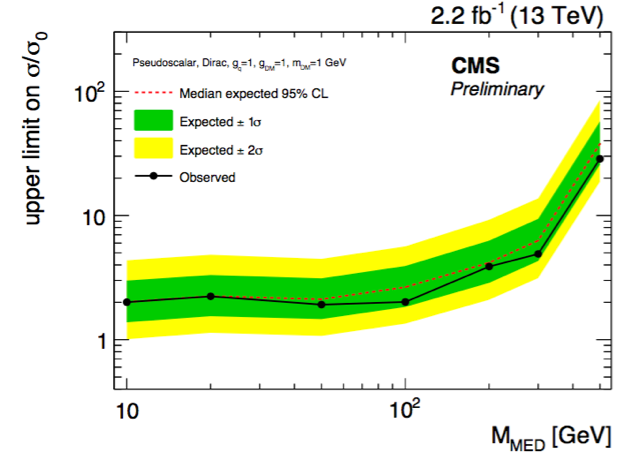}
\caption{Exclusion limits at 95\% C.L. for the scalar (left) and pseudo-scalar (right) resonances for the ATLAS\cite{ATLAS-CONF-2016-050} and CMS\cite{CMS-EXO-16-005} results, respectively.}
\label{figttmet}
\end{figure}

For the ATLAS collaboration, the searches for the associated production of a top quark pair and a DM pair are performed per decay channel : first in the single lepton channel \cite{ATLAS-CONF-2016-050} but also in the dilepton  \cite{ATLAS-CONF-2016-076} and fully hadronic \cite{ATLAS-CONF-2016-077} channels. In each analysis, a similar strategy is followed. It uses 2015 and 2016 datasets (amounting to 13.3 fb$^{-1}$) and are based first on the definitions of signal regions, based on discriminating (signal specific) variables. Then, several control regions, enriched in specific types of backgrounds, are used to control the backgrounds from data. Finally, validation regions are used to validate the background estimations. Counting experiments are used. No evidence of signal is found and the corresponding exclusion limits  at 95\% Confidence Level (C.L.) are presented on the left hand of the Fig.\ref{figttmet} for the scalar mediator case.

For the CMS collaboration, the search is a combination of the semi-leptonic and fully hadronic channels analyses, in order to reach the best possible sensitivity \cite{CMS-EXO-16-005}. In the full hadronic channel, a categorization of events is done based on the number of reconstructed hadronic tops per event. For each channel and each category, $\slashed{E}_T$ distributions are used to extract the signal. To estimate the background contaminations, various control regions, enriched in specific backgrounds, are used simultaneously in a fit. The analysis uses 2.2 fb$^{-1}$ of the 2015 dataset.   No evidence of signal is found and the corresponding exclusion limit  at 95\% C.L. is presented on the right hand of the Fig.\ref{figttmet} for the pseudo scalar mediator case. Limits are also provided for all the benchmark points suggested by the DM forum \cite{DMforum}.

\section{Search for DM with monotop signatures}
\label{sec:monotop}

\begin{figure}[!htb]
\centering
\includegraphics[height=1.9in]{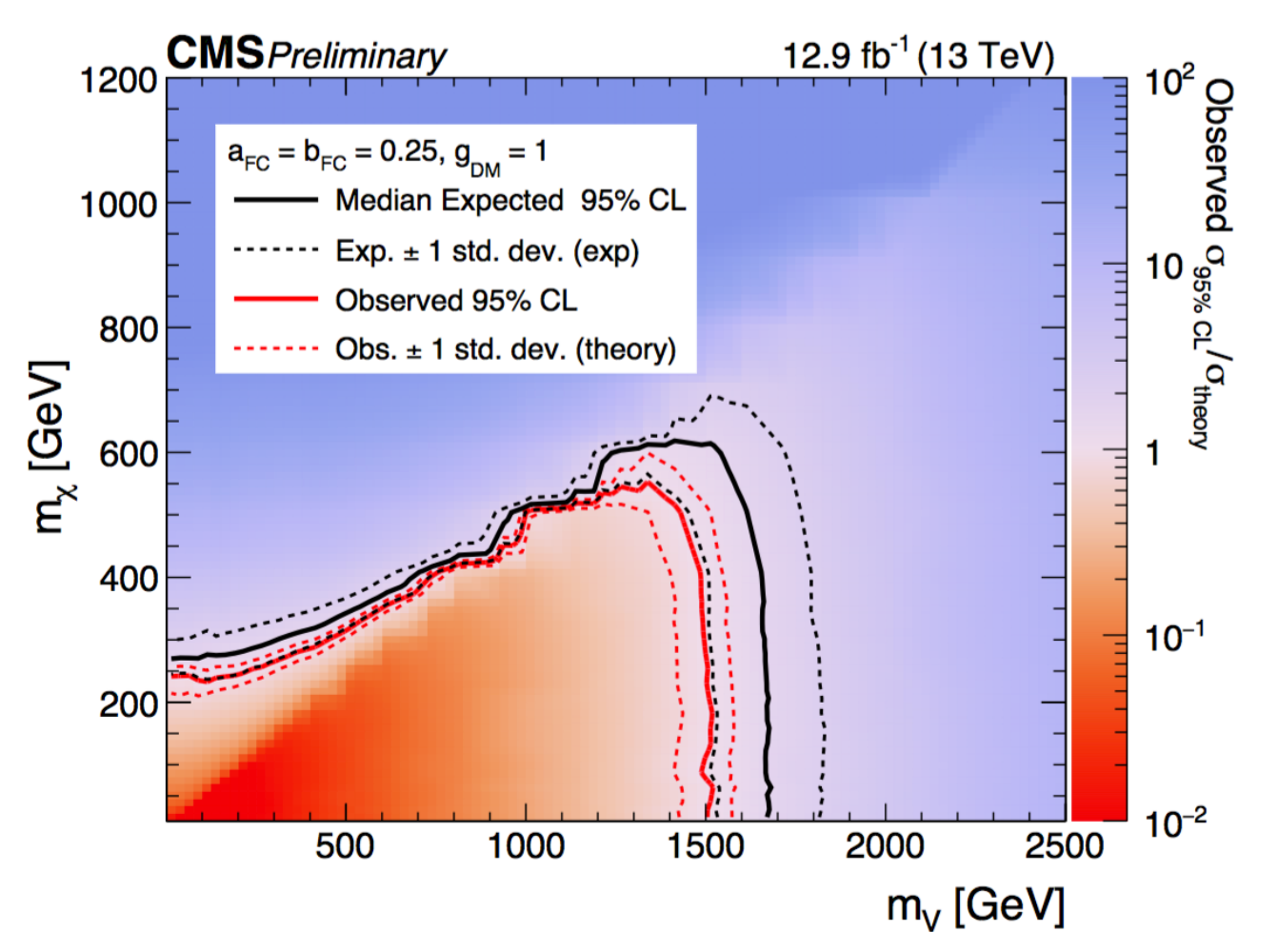}
\includegraphics[height=1.9in]{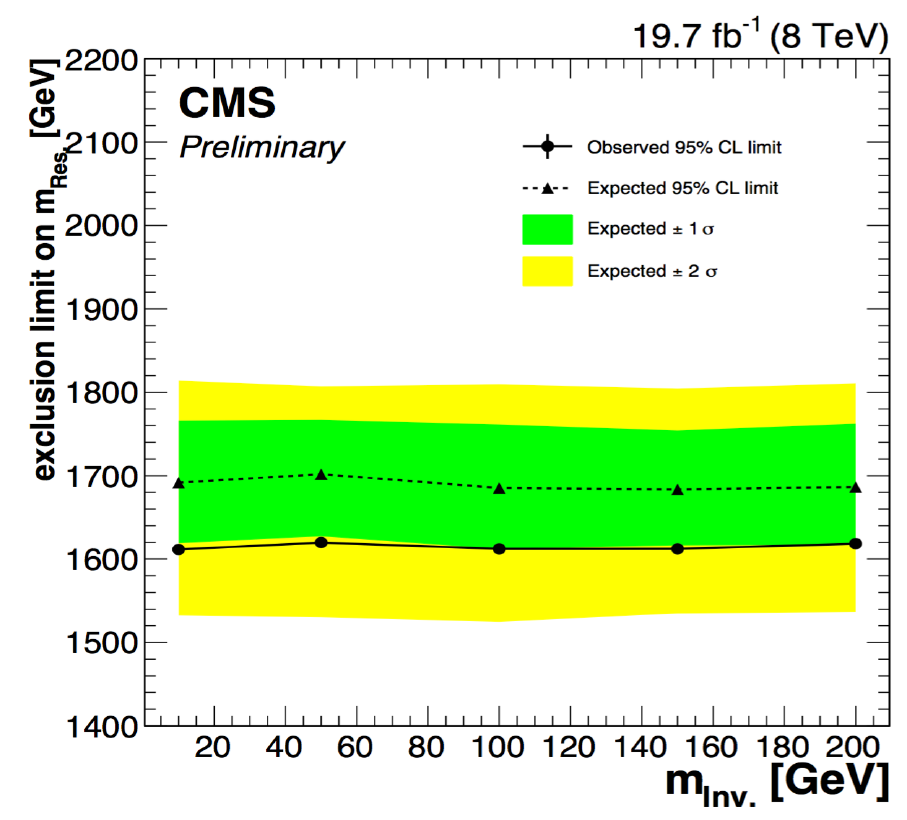}
\includegraphics[height=1.9in]{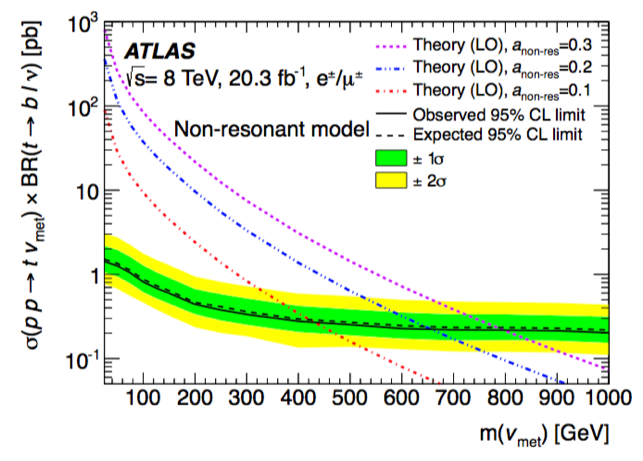}
\caption{Exclusion limits at 95\% C.L. for the resonant scenario in the hadronic  (top left, for $a_{res}$ and $b_{res}=1$) \cite{CMS-EXO-16-040} and the muonic (top right) \cite{CMS-B2G-15-001} channels from CMS, and for the non-resonant scenario in the leptonic channel (bottom) from ATLAS \cite{ATLASmonotop}.}
\label{figmonotop}
\end{figure}

In the CMS collaboration, the searches for monotop are performed either in the hadronic or in the muonic channels. In the hadronic channel the most recent CMS result \cite{CMS-EXO-16-040}, which uses about 13 fb$^{-1}$  of data, the signal is searched for by means of a fit of the $\slashed{E}_T$ distribution for boosted top quarks topologies.  A simultaneous fit of the signal and control regions is performed. No evidence of signal is found and the corresponding exclusion limits  at 95\% C.L. are presented on the top-left plot of the Fig.\ref{figmonotop} for the resonant scenario. A search is also performed in the muonic channel at 8 TeV using about 20 fb$^{-1}$  of data \cite{CMS-B2G-15-001}. The search is done by performing a simultaneous fit of signal and control regions using the transverse mass of the $W$ boson as the discriminating variable. No evidence of signal is found and the corresponding exclusion limits  at 95\% C.L. are presented on the top-right plot of  Fig.\ref{figmonotop} for the resonant scenario.

The ATLAS collaboration studied monotop production in the leptonic channel (electron and muon) using  about 20 fb$^{-1}$  of 8 TeV data \cite{ATLASmonotop}. The analysis is a counting experiment using a signal region and control regions to extract the background contamination. No evidence of signal is found and the corresponding exclusion limits  at 95\% C.L. are presented on bottom plot of Fig.\ref{figmonotop} for the non resonant scenario.

\section{Conclusion}
\label{sec:conclusion}

Top signatures provide very interesting probes for searching for Dark Matter. For the search of the associated production of top quarks and DM particles, two approaches are followed : the production of top-pairs or the single top production.
In both cases, the signals are described by simplified models, including new scalar (pseudo-scalar) resonances or flavor changing interactions. A global effort for harmonizing the theoretical conventions has been performed within the Dark Matter Forum. Searches for $t\bar{t}+\slashed{E}_T$ and monotop have been performed at 8 and 13 TeV and no signs of new physics have been observed so far.

%%%%%%%%%%%%%%%%%%%%%%%%%%%%%%%%%%%%%%%%%%%%%%%%%%%%%%%%%%%%%%%%%%%%%%%%%
%%
%%   use this format to include an .eps figure into your paper
%%

%%%%%%%%%%%%%%%%%%%%%%%%%%%%%%%%%%%%%%%%%%%%%%%%%%%%%%%%%%%%%%%%%%%%%%%%%%%

%\Acknowledgements


\begin{thebibliography}{99}

%%
%%  bibliographic items can be constructed using the LaTeX format in SPIRES:
%%    see    http://www.slac.stanford.edu/spires/hep/latex.html
%%  SPIRES will also supply the CITATION line information; please include it.
%%

\bibitem{DMforum} Dark Matter Benchmark Models for Early LHC Run-2 Searches: Report of the ATLAS/CMS Dark Matter Forum,  arXiv:1507.00966.
\bibitem{atlas} The ATLAS experiment at the CERN LHC, JINST 3:S08003,2008.
\bibitem{cms} The CMS experiment at the CERN LHC, JINST 3:S08004, 2008.
\bibitem{ATLAS-CONF-2016-050} Search for top squarks in final states with one isolated lepton, jets, and missing transverse momentum in $\sqrt{s}$ = 13 TeV pp collisions with the ATLAS detector, ATLAS-CONF-2016-050.
\bibitem{ATLAS-CONF-2016-076} Search for direct top squark pair production and Dark Matter production in final states with two leptons in $\sqrt{s}$ = 13 TeV pp collisions using 13 fb$^{?1}$ of ATLAS data, ATLAS-CONF-2016-076.
\bibitem{ATLAS-CONF-2016-077} Search for the Supersymmetric Partner of the Top Quark in the Jets+Emiss Final State at $\sqrt{s}$ = 13 TeV, ATLAS-CONF-2016-077.
\bibitem{CMS-EXO-16-005} Search for dark matter in association with a top quark pair at $\sqrt{s}$=13 TeV, CMS-EXO-16-005.
\bibitem{CMS-EXO-16-040} Search for new physics in a boosted hadronic monotop final state using 12.9 fb$^{?1}$ of $\sqrt{s}$=13 TeV data, CMS-EXO-16-040.
\bibitem{CMS-B2G-15-001} Search for monotop in the muon channel in proton-proton collisions at $\sqrt{s}$= 8 TeV,  CMS-B2G-15-001.
\bibitem{ATLASmonotop} Search for invisible particles produced in association with single-top-quarks in proton-proton collisions at $\sqrt{s}$ = 8 TeV with the ATLAS detector, Eur. Phys. J. C (2015) 75:79
 
 
\end{thebibliography}
\end{document}